\documentstyle[preprint,prb,aps,psfig]{revtex}

\begin{document}
\tightenlines

\title{Influence of spin fluctuations on the superconducting transition
temperature and resistivity in the $t-J$ model at large N}

\author{{A. Greco}$^{a,b}$ and {R. Zeyher}$^a$} 

\address{$^a$ Max-Planck-Institut\  f\"ur\
Festk\"orperforschung,\\ Heisenbergstr.1, 70569 Stuttgart, Germany \\
$^b$Permanent address: Departamento de F\'{\i}sica, Facultad de
Ciencias Exactas e Ingenier\'{\i}a and \\
IFIR(UNR-CONICET), Av. Pellegrini 250, 
2000-Rosario, Argentina}

\date{\today}

\vspace{5cm}

\maketitle

\begin{abstract} 

Spin fluctuations enter the calculation of the superconducting transition 
temperature $T_c$ only in the next-to-leading order
(i.e., in $O(1/N^2)$ of the $1/N$ expansion of the $t-J$ model. We have 
calculated these terms and
show  that they have only little influence on the value for $T_c$ obtained
in the leading order $O(1/N)$ in the optimal and overdoped region, i.e., for 
dopings larger than the instability towards a flux phase. 
This result disagrees with recent spin-fluctuation mediated pairing 
theories. The discrepancies can be traced back to the fact that in our
case the coupling between electrons and spins is 
determined by the t-J model and not adjusted and that the spin
susceptibility is rather broad and structureless 
and not strongly peaked at low energies as in spin-fluctuation models.

Relating $T_c$ and transport we show that the effective interactions in the 
particle -particle and particle-hole channels are not simply related
within the 1/N expansion
by different Fermi surface averages of the same interaction  as in the case
of phonons or spin fluctuations. 
As a result, we find that large values for $T_c$ and rather small scattering 
rates in the normal state as found in the experiments can easily be
reconciled with each other. We also show that correlation effects 
heavily suppress transport relaxation rates relative to quasiparticle
relaxation rates in the case of phonons but not in the case of spin 
fluctuations.

\end{abstract}

\newpage
\section{Introduction}

Calculations of the superconducting transition temperature $T_c$ and other
specific properties of the cuprates are usually based on the Hubbard model.
One approach assumes that the effective Hubbard repulsion $U$ is smaller or
comparable to the electronic hopping ampliutde $t$ so that approximations
which become exact at weak coupling are taken as a basis for 
calculations of properties of the cuprates. The FLEX 
approximation\cite{Bickers},
which sums bubble and ladder diagrams in a self-consistent way,
is such an approximation which has been rather successful in reproducing
properties of the cuprates\cite{Grabowski}. 
A rather different and certainly more
realistic view of the cuprates assumes that $U$ is substanially larger
than $t$ at all relevant dopings. As a result the strong-coupling limit
of the Hubbard model applies which 
is equivalent, except for three-center potentials, to the $t-J$ model.
Cuprates are then no longer treated as moderately correlated metals 
as in the FLEX approximation but
as doped Mott insulators. Formally this means that self-energies, effective
interactions, etc., are not expanded in powers of $U$ but in $1/U$
and that, as a new point, the constraint of having no double occupancies of 
sites has to be taken into account. Throughout our paper we adopt this 
second point of view and thus assume that the strong-coupling limit applies
to the cuprates over the whole doping region including also the
overdoped case.

The $t-J$ model has no small parameter which would allow a systematic
perturbation expansion. One way out is to extend this model in such
a way that a small parameter arises and that the properties of the original
model are obtained as power expansion in this small paramter.
One convenient generalization consists of replacing the
two spin components of the $t-J$ model by $N$ components and, correspondingly,
its orginal $SU(2)$ symmetry by the symplectic symmetry group $Sp(N/2)$. In
this way properties of the physical $N=2$ model can be expanded around the
$N=\infty$ limit of the extended model which describes renormalized, but
non-interacting, particles. The interaction between the particles can then be
taken into account by means of a systematic $1/N$ expansion. We have 
formulated this expansion directly in terms of $X$-operators
which respect the constraint via their commutator
and anticommutator rules. Applying this method to
superconducting instabilities we found \cite{Greco1,Zeyher1,Zeyher2} 
for a square lattice relevant 
instabilities only in the d-wave channel ($\Gamma_3$ representation of the 
point group $C_{4v}$ of the square lattice) due to the Heisenberg
term. Putting $N=2$, the corresponding
transition temperature $T_c$ was about 0.03 in units of the nearest-neighbor
hopping element $t$ for a doping value away from half-filling 
of $\delta = 0.17$.

The above calculation took into account only the leading order of the
$1/N$ expansion. Thus the anomalous self-energy was calculated in 
$O(1/N)$, or equivalently, the kernel $\Theta$ of the linearized gap 
equation in $O(1)$, since an overall factor $1/N$ has been taken out
in the gap equation
(see Eq.(40) of Ref.\cite{Zeyher2}, or Eq.(1) below). Although this order 
contains also spin-flip contributions it does not contain
the well-known RPA terms for the spin susceptibility. These terms are,
of course, present in the $1/N$ expansion but their leading contribution
(one bubble) yields only $1/N$ terms to $\Theta$ and thus has been
dropped in Ref.\cite{Zeyher2}. 

Presently many experimental data
are interpreted as giving evidence for a spin-fluctuation mediated
pairing mechanism for high-T$_c$ superconductivity 
\cite{Pines1,Pines2,Pines3,Carbotte}. It is thus interesting
to include the above terms, at least, if they are of $O(1/N)$ in 
$\Theta$, in the calculation of T$_c$. 
It is the aim of the present communication to do this
and to consider their effects on T$_c$ and the resistivity.
Using the $1/N$ expansion of the $t-J$ model these terms
can easily be identified. The corresponding coupling strength
between electrons and spins is determined uniquely by the model 
parameters and has not to be
chosen in an ad hoc way. At the same time such a calculation yields a check
for the validity of the $1/N$ expansion for T$_c$: If, for instance,
the next-to-leading terms are of the same magnitude as the leading ones,
after putting $N=2$, a rapid convergence of the series cannot be expected.
Of course, the comparison of the two orders would be most convincing if all 
the $O(1/N)$ terms of $\Theta$
could be taken into account. The resulting expressions are, however, rather
involved and their evaluation has to be postponed. In the following we will 
restrict
ourselves to that part of $\Theta$, called $\Theta_s$, which is of $O(1/N)$ 
and describes RPA-like spin fluctuations. These terms are derived in
section 2. Section 3 discusses properties of the resulting spin susceptibilty
and compares them with a different theoretical approach and also with 
experimental data. In section 4
properties of $\Theta_s$, which are relevant for superconductivity,
are discussed and their influence on T$_c$ is calculated. In section 5
the consequences of $\Theta$ for transport properties are investigated.
Section 6 contains our conclusions. Throughout the paper we will restrict
ourselves to dopings $\delta > \delta_c$ where $\delta_c$ denotes the
critical value for the instability towards a flux or bond-order phase     
\cite{Zeyher1,Zeyher2,Morse,Cappelluti1}. $\delta_c$ is about 0.15 for 
$J/t=0.3$.
It has been argued\cite{Cappelluti1} that $\delta_c$ corresponds to optimal
doping so that we deal in the following with optimal doping and the overdoped
regime.

\section{Basic equations}

Using a $1/N$ expansion for the $t-J$ model the linearized equation
for the superconducting gap $\Sigma_{an}$ can be written as \cite{Zeyher2}
\begin{eqnarray}
{\Sigma}_{an}(k) = -{T \over {NN_c}}\sum_{k'}\Theta(k,k')
{1 \over {\omega_{n'}^2 +\epsilon^2({\bf k'})}}
{\Sigma}_{an}(k').
\label{sup} 
\end{eqnarray}
$N$ denotes the number of electronic degrees of freedom per site.
We assume that they consist of $N/2$ replica of one orbital with two spin
components and $N$ can be assumed to be an even integer for convenience.
$N_c$ is the number 
of primitive cells and $k$ the supervector $k=(\omega_n,{\bf k})$,
where $\omega_n$ denotes the fermionic Matsubara frequency $\omega_n =
(2n+1)\pi T$, $\bf k$ the wave vector, and $T$ the temperature.
$\epsilon({\bf k})$ is the one-particle energy 
\begin{equation}
\epsilon({\bf k}) = {\delta \over 2}t({\bf k})-J({\bf k}) \cdot
{1 \over N_c} \sum_{\bf p} cos(p_x)f(\epsilon({\bf p}-\mu)),
\label{energies}
\end{equation}
with
\begin{equation}
t({\bf k}) = -2|t|(cos(k_x) +cos(k_y)),
\label{t}
\end{equation}
\begin{equation}
J({\bf k}) = 2J(cos(k_x)+cos(k_y)).
\label{J}
\end{equation}
$\mu$ is the chemical potential, $f$ the Fermi function,
$\delta$ the doping away from half 
filling and the lattice constant of the square lattice has been put to
one.
Note also that the usual definition of the hopping matrix element in $N=2$
theories corresponds to $|t|/2$ in our notation. Generally accepted values for
$|t|/2$ and $J$ are about $0.4 eV$ and $0.15 eV$, respectively.
In the following we will use $|t|/2$ as energy unit. The kernel $\Theta$
represents the effective interaction in the particle-particle channel. Its 
leading $O(1)$ contributions consist of
four different terms $\Theta^{(1)}$, $\Theta^{(2)}$, $\Theta^{(3)}$, and
$\Theta^{(4)}$, given explicitly in Eqs.(42)-(52) of Ref. \cite{Zeyher2}.
The first one describes the instantaneous part, the second one fluctuations
in the charge density, and the third and fourth ones mixed spin and charge
fluctuations due to the anomalous part of the vertex. 

$\Theta^{(2)}$ originates from single  particle and collective 
electron-hole exitations in the
charge channel. A similar contribution $\Theta_s$ is also obtained in the 
spin channel
\begin{equation}
\Theta_s(k,k') = -(t({\bf k'})+J({\bf k}-{\bf k'}))\gamma_s(k,k'-k),
\label{spin}
\end{equation}
where the sign for singlet pairing has been chosen.
Expanding $\gamma_s$ in powers of $1/N$ the leading $O(1)$ contribution 
is just a product of delta-functions. This contribution has already been taken
into account in $\Theta^{(1)}$. The $O(1/N)$ contribution of $\gamma_s$
contains the 
term
\begin{equation}
\gamma_s'(k_1,k_2)={1 \over N} (t({\bf k}_1)+J({\bf k}_2))\chi^{(0)}(k_2),
\label{free}
\end{equation}
with
\begin{equation}
\chi^{(0)}(k) = {1 \over N_c} \sum_{\bf q} {
{f(\epsilon({\bf k}+{\bf q}))-f(\epsilon({\bf q}))} \over
{\epsilon({\bf k} + {\bf q}) -\epsilon({\bf k}) - i\omega_n}}.
\label{susfree}
\end{equation}
Inserting Eq.(\ref{free}) into Eq.(\ref{spin}) we have the following
additional $O(1/N)$ contribution to the kernel $\Theta_s$ due to spin 
fluctuations 
\begin{equation}
\Theta_s(k_1,k_2) = -{1 \over N}(t({\bf k}_2)+J({\bf k}_1-{\bf k}_2))
(t({\bf k}_1)+J({\bf k}_1-{\bf k}_2))\chi^{(0)}(k_1-k_2).
\label{theta}
\end{equation}

Both the susceptibility $\chi^{(0)}$ and the coupling functions on the 
right-hand side of Eq.(\ref{theta}) are different from what one may expect
from a simple spin-fluctuation mediated theory for superconductivity.
The coupling functions (first and second factors in the above expression)
contain not only the potential $J$ but also the kinetic parameter $t$.
The origin of the latter lies in the fact that fermionic Hubbard operators
are not usual creation and annihilation operators and generate many-body
interactions in the perturbation expansion. 
Secondly, one would expect the RPA expression for the spin susceptibility 
on the right-hand side of Eq.(\ref{theta}) and not the equivalent of
just one bubble, as described by Eq.(\ref{susfree}). The reason for this 
is that in RPA-type diagrams each bubble is multiplied by $J/N$, i.e., 
entering as an $O(1/N)$
contribution. Thus the infinite summation over bubbles in the RPA is not
in the spirit of the 1/N expansion: Before the product of even two bubbles 
can be included one has to sum up infinite series of diagrams
of different types. This can be seen from the O(1/N) expression for the
spin vertex $\gamma_s$ in Refs. \cite{Kulic,Cappelluti2}. Since $\chi^{(0)}$
already exhausts more than half of the exact sum rule for the spin
susceptibility (see below) we retain from the O(1/N) terms to $\Theta$ only
the term $\Theta_s$, as given by Eq.(\ref{theta}).

\section{Sum rule and properties of the spin susceptibility}

For a general $N$ the spin operator at site $i$ is replaced by
\begin{equation}
S_i^\alpha \rightarrow  S_i^\alpha(\nu),
\label{def}
\end{equation}
where $\nu$ is a flavor index counting the $N/2$ copies of the original
orbital and $\alpha$ a Cartesian index counting
the three vector components of the spin. Expressed by Hubbard operators 
the z-component of the spin operator acting on the copy $\nu$ reads
\begin{equation}
S^z_i(\nu) = {1\over 2}(X_i^{2\nu,2\nu}-X_i^{1\nu,1\nu}).
\label{sz}
\end{equation}
The spin susceptibility $\chi^{\alpha\beta}$ is defined in Fourier
space for a general $N$ by
\begin{equation}
\chi^{\alpha\beta}({\bf k},\omega) = {2 \over N}\; (-i)\sum_{\nu=1...N/2}
\int_0^\infty dt e^{i\omega t} \langle [S^\alpha_{\bf k}(t,\nu),
S^\beta_{-{\bf k}}(0,\nu)] \rangle,
\label{chim}
\end{equation}
where $\langle..\rangle$ denotes a thermodynamic average  
and $[...]$ the commutator. 
Inserting Eq.(\ref{sz}) into Eq.(\ref{chim}), using the 
fluctuation-dissipation theorem, summing over $\bf k$ and integrating over 
$\omega$
we obtain for the z-component of the imaginary part of $\chi$ in the normal 
state
\begin{equation}
-{1\over N_c} \sum_{\bf k} \int^\infty_{-\infty} Im\chi^{zz}({\bf k},
\omega)(1+n(\omega)) = {\pi \over {2N}} \sum_{\nu=1...N/2} \langle 
(X_i^{2\nu,2\nu}-X_i^{1\nu,1\nu})^2 \rangle.
\label{sum}
\end{equation}
$n(\omega)$ is the Bose function. The right-hand side of 
Eq.(\ref{sum}) can be evaluated exactly for the two cases $N=2$ and
$N=\infty$. For $N=2$, which corresponds to the physical case
of one orbital with two spin components, the $X$-operators are projection
operators, so that Eq.(\ref{sum}) becomes,
using also the constraint $X_i^{00}+X_i^{11}+X_i^{22}=1$,
\begin{equation}
-{1\over N_c} \sum_{\bf k} \int^\infty_{-\infty} d \omega 
Im\chi^{zz}({\bf k},
\omega)(1+n(\omega)) = {\pi \over 4} (1-\delta),
\label{sumrule}
\end{equation}
where $\delta$ is the doping away from half-filling.
Eq.(\ref{sumrule}) is exact and generalizes the spin sum rule
to a general doping $\delta$. 
Experimental data are usually given in terms of the magnetic susceptibiltiy
which is the product of the spin susceptibility and $g^2\mu_B^2$, where
$g$ and $\mu_B$ are the gyromagnetic ratio and the Bohr magneton,
respectively. Following a widely used notation we denote the 
negative imaginary part of the magnetic susceptibility,
analytically continued to the real axis 
$i\omega_n \rightarrow \omega + i \eta$, by $\chi''({\bf k},
\omega)$. Inserting $g=2$ the sum rule Eq.(\ref{sumrule}) assumes then
the form
\begin{equation}
{1\over N_c} \sum_{\bf k} \int_{-\infty}^\infty d \omega 
\chi''({\bf k},\omega)(1+n(\omega)) = \mu_B^2 \pi (1-\delta).
\label{sumrule2}
\end{equation}

For $N>2$ the projection properties of the $X$ operators are lost and
non-diagonal averages on the right-hand side of Eq.(\ref{sum}) are nonzero.
However, at large $N$'s the non-diagonal averages are by a factor $1/N$
smaller than the diagonal ones and thus can be omitted. The diagonal ones
can be calculated using Eq.(\ref{susfree}) and one obtains intead of
Eq.(\ref{sumrule2})
\begin{equation}
{1\over N_c} \sum_{\bf k} \int_{-\infty}^\infty d \omega 
\chi''({\bf k},\omega)(1+n(\omega)) = \mu_B^2 \pi (1-{\delta}^2)/2.
\label{sumrule3}
\end{equation}
At large dopings Eqs.(\ref{sumrule2}) and (\ref{sumrule3}) agree
approximately but at zero doping the right-hand side
of Eq.(\ref{sumrule3}) is only $1/2$ of that of Eq.(\ref{sumrule2}).
The right-hand side of the above sum rules, considered as a function of
$N$, interpolates between the above two extreme cases $N=2$ and $N=\infty$. 
This means that terms $\sim 1/N$ and of higher orders contribute to
the sum rule and, if evaluated at $N=2$, cancel the factor $(1+\delta)/2$
on the right-hand side of Eq.(\ref{sumrule3}). This implies, in particular,
that the sum rule for the spin susceptibility depends on the strength of the
interaction between the particles.

Fig. 1 shows theoretical and experimental data for the momentum averaged
magnetic susceptibility $\chi''(\omega)$, defined by
\begin{equation}
\chi''(\omega)= {1 \over N_c} \sum_{\bf k} \chi''({\bf k},\omega).
\label{inte}
\end{equation}
The dot-dashed line represents the susceptibility at 
large $N$'s, $\chi_{\infty}''(\omega)$.
The dashed curve in Fig. 1 is a plot of the function 
$\chi_P''(\omega)$ which is associated with the following phenomenological 
expression for the spin susceptibility in high-T$_c$ compounds 
\cite{Pines1,Pines2,Pines3},
\begin{equation}
\chi_P({\bf k},\omega) = {{\chi_Q} \over{-i \omega/ \omega_{sf} 
+(1 +\xi^2({\bf k}-{\bf Q})^2)}}.
\label{Pines}
\end{equation}
Here we have used the values listed in Table II of Ref. \cite{Pines3}
for the various parameters for $\delta=0.2$ (second-last column in Table II). 
The open circles in Fig. 1 are experimental points for the momentum
averaged susceptibility divided by 2 in underdoped 
$YBaCu_3O_{6.6}$\cite{Hayden} at $T=80K$.
Analogous experimental data
have been presented in Refs.\cite{Bourges1,Bourges2,Fong} for various
dopings in $YBa_2Cu_3O_{7-x}$. The data for $YBa_2Cu_3O_{6.7}$
(Fig. 6 of Ref.\cite{Fong}), covering the
energy window between 0 and 70 meV, are smaller by roughly a factor 2,
but otherwise rather similar: They also show a first maximum at around 30 meV
and a  second maximum, which lies at a somewhat lower energy of about 60 meV 
and is broader and less pronounced compared to the experimental data shown in 
Fig. 1.
Increasing the doping in $YBa_2Cu_3O_{7-x}$ spin scattering
in the normal state becomes weaker and finally undetectable in the optimally 
doped case\cite{Fong}. $\chi_\infty''$
exhausts a little more than half of the sum rule Eq.(\ref{sum}). 
It spreads in frequency
over the bandwidth of about 0.7 eV and exhibits a broad 
maximum near 0.3 eV. Integrating $\chi_\infty''(1+n(\omega))$ up to the
largest measured frequency the dot-dashed and the experimental curve
exhaust about 1/6 of the sum rule\cite{Aeppli}.  However, the experimental 
points
for the oxygen contents 6.6 and 6.7 show a dramatic shift of spectral weight 
from high to low frequencies which is absent in $\chi_\infty''$.
The dashed curve in Fig. 1 is of similar magnitude 
as the experimental curve between 0 and about 70 meV. However,
it misses the strong decrease of the experimental curve above
70 meV and is therefore much too large at high frequencies.

Fig. 2 shows data for $\chi''({\bf Q},\omega)$ at ${\bf Q}=(\pi,\pi)$
where the susceptibilities assume either their maximal values in ${\bf k}$-
space or are near their maximal values in case where a splitting into
incommensurate peaks positioned near $\bf Q$ occurs. The filled circles are
experimental values from Ref.\cite{Fong} for $YBa_2Cu_3O_{6.7}$ at $T=200K$.
The dot-dashed and dashed curves in Fig. 2 represent 
$\chi''_\infty({\bf Q},\omega)$, multiplied by 10, and 
$\chi''_P({\bf Q},\omega)$, respectively. 
Note that there is a scale change of about a factor 50 along the
y-axis between Figs. 1 and 2. This means that the experimental spectral
weight of the spin susceptibility is strongly concentrated near $\bf Q$
and at low energies around 20 or 30 meV. This strong localization in momentum
and energy is absent to a large extent in $\chi''_\infty$ and to a minor
degree also in $\chi''_P$. Plotted as a function of momentum for a fixed
frequency $\chi_\infty''$ shows well-pronounced incommensurate peaks
near $\bf Q$ in spite of the fact that 
$\chi_\infty''({\bf Q},\omega)$ as a function of $\omega$ is according to 
Fig. 2 rather small and structureless. 

The showing of experimental points for the underdoped regime in 
Figs. 1 and 2 does not mean that $\chi''_{\infty}$ should reproduce these 
experimental spectra. As mentionned in the introduction our $1/N$ expansion
holds in this form only for dopings $\delta$ larger than $\delta_c$
where the instability towards a flux phase occurs, i.e., as discussed in 
Ref.\cite{Cappelluti1}, in the optimally doped  and overdoped region.
Presently spin scattering in optimally doped $YBa_2Cu_3O_7$ is too
weak to be detectable by inelastic neutron scattering with the 
instrumentation currently available\cite{Fong}. Recently, 
an upper limit for $\chi''({\bf Q},\omega)$ of $45 \mu_B^2/eV$ has been given
for the instrumentally accessible energy range $20 meV < \omega < 50 meV$
\cite{Sidis} which is substantially lower than the dashed line in Fig. 2. 
This suggests that a low-frequency peak in $\chi''$ is present only at low 
dopings and that
$\chi''$ in optimally doped and overdoped samples extends over
a large energy region because of the exact sum rule Eq.(\ref{sumrule}). 
One thus may argue that 
$\chi''_\infty$ may represent well the spin susceptibility in the optimal and 
overdoped
regimes and that the experimental points as well as $\chi''_P$ 
in Figs.1 and 2 only apply to the underdoped case.

Another quantity characterizing $\chi$ is the coherence length
$\xi$. In an isotropic two-dimensional description the decay of the
the real part of $\chi({\bf r},\omega=0)$ in real space is given by 
$K_0(|{\bf r}|/\xi)$, where $K_0$ is a modified Bessel function of
zeroth order. The dashed and dot-dashed lines in Fig. 3 represent this 
Bessel function for
$\xi=2.3$ and $\xi=1.0$, respectively. The diamonds are the absolute values 
for the real part using $\chi_P$ at various neighbor sites of the square 
lattice. The diamonds lie perfectly on the Bessel 
function with $\xi=2.3 $ as expected. The circles in Fig. 3 
have been calculated with the large-N susceptibility 
$\chi_\infty$.
Because of the anisotropy of the lattice not all of the calculated points 
lie on the curve for the Bessel function. Fig. 3 nevertheless shows that the 
corresponding $\xi$ is near the value 1. Repeating the same calculation
at finite temperatures we find that $\xi$ varies with temperature
on the scale of $t$ and thus is independent of temperature 
in the range of experimental interesting temperatures. Experimental values 
for $\xi$ from inelastic neutron scattering are independent of temperature
and have been summarized in Fig. 2
of Ref.\cite{Balatsky}as a function of doping. For optimally doped 
$YBa_2Cu_3O_{7-x}$
$\xi$ is only little larger than 1 which also agrees with NMR data
\cite{Gorny,Berthier}. This supports the idea that $\chi_\infty$ represents
a sensible approximation for the spin susceptibility in 
the optimal and overdoped regime. 

\section{Properties of $\Theta_s$ near the Fermi surface and its influence
on the transition temperature $T_c$}

A relevant quantitiy for superconductivity is $\Theta({\bf k},{\bf k'},
\omega=0)$ with momenta ${\bf k}$ and ${\bf k'}$ put to the Fermi surface.
Fig. 4 shows this quantitiy for a fixed first momentum at 
${\bf k}= (2.465,0.309)$ as a function of the second momentum $\bf k'$.
${\bf k'}$ moves counterclockwise around the Fermi line passing through the
points $X,Y,(\bar{X},\bar{Y})$ along the positive (negative) x- and y-axis,
respectively. The dashed line corresponds to the
kernel $g^2\chi_P/8$, using the coupling constant $g=0.6 eV$ from Table II
of Ref. \cite{Pines3}. The dot-dashed and solid lines in Fig. 4 describe
the kernel $\Theta_s$  and the full O(1) kernel
$\Theta$ calculated in Ref.\cite{Zeyher2}, respectively. The doping is
always $\delta=0.17$ and $J$ equal to 0.3. The additional factor $1/N$
in $\Theta_s$ has been taken into account using the physical value $N=2$.
The curves are not symmetric with respect of $X$ or $\bar{X}$ because
the first argument ${\bf k}$ has been put to a general point somewhat away
from $X$.

First we note that all three
curves have maxima at $Y$ and $\bar{Y}$, i.e., for momentum transfers
of ${\bf Q}=(\pi,\pi)$ or of $-{\bf Q}$. The dashed and dot-dashed 
curves correspond
to purely repulsive, the solid one to predominantly repulsive interactions.
As a result d-wave pairing is preferred in all three cases. The variation
in the solid curve along the Fermi surface, which for d-wave pairing is
the relevant quantitiy, is about two to three times larger than for the
dot-dashed curve. This means that the spin-fluctuation mediated interaction
is by this factor smaller compared to the sum of all O(1) contributions
to the effective interaction. This already indicates that the 1/N expansion
makes sense insofar that the terms decrease with increasing order in $1/N$.
The dot-dashed line in Fig. 4 exhibits some fine structure near the point 
$\bar {Y}$ which is caused by the splitting of the peak at ${\bf Q}$ into
incommensurate peaks displaced slightly away from $\bf Q$.
The appearance of these incommensurate peaks due to the Fermi-surface
geometry has been discussed previously in Ref. \cite{Levin1}. 

The kernel $\Theta({\bf k},{\bf k'},0)$ represents the effective
interaction in the particle-particle channel at zero frequency. The solid
line in Fig. 4 thus can be compared with the upper curve in Fig. 8
of Ref. \cite{Bulut} which has been obtained from Monte Carlo simulations.
Though both kind of curves show minima at the X and maximma at the Y points
the energy scale of the curve in Ref. \cite{Bulut} is one order of magnitude
larger than in our case and also has a large momentum independent part.
These differences can roughly be explained by the fact that the energy
scale in Ref. \cite{Bulut} is set by the Hubbard constant $U$ and in our
case by $J({\bf k})$. The parameter values used in Ref. \cite{Bulut} thus 
correspond more to the weak-coupling case 
whereas we deal with strong-coupling. Fig. 4 also shows that
the momentum dependence of the effective interaction is similar to that
of the spin susceptibility which determines the momentum dependence of
the dashed and dot-dashed curves. However, one cannot conclude from this
that the effective interaction is caused by antiferromagnetic spin
fluctuations because the solid line does not include them. 

The dashed curve in Fig. 4 is at least one order of magnitude larger than 
the dot-dashed
curve, especially, near the point ${Y}$. Though both terms describe
interactions due to spin fluctuations their strengths differ roughly by one
order of magnitude. 
To understand why these curves are so different in magnitude
we characterize the coupling in the large-N expression
Eq.(\ref{theta}) by an effective coupling constant $\bar{g}$. Since both, 
$t$ and $J$, depend on the momentum we define $\bar{g}$ by the following
average value appropriate for d-wave symmetry
\begin{equation}
\bar{g}^2 = << \Theta_s({\bf k},{\bf k'},\omega=0)>>_3/
<<\chi^{(0)}({\bf k},{\bf k'},\omega=0)>>_3.
\label{coupling}
\end{equation}
$<<...>>_3$ denotes in the above equation a momentum average over the Fermi 
line
with a weight function with d-wave symmetry. An appropriate weight function
is given by the eigenvector associated with the lowest eigenvalue with
d-wave symmetry. The explicit form of the corresponding
kernel yielding the desired eigenvalues and eigenvectors has been described
in Ref.\cite{Greco1}. This definition can be extended to any 
momentum-dependent operator.
The Fermi surface average is then obtained as the lowest eigenvalue
of the associated operator. The dot-dashed and the 
dotted lines in Fig. 5 show
the numerator and the denominator of the right-hand side of Eq.(\ref{coupling})
as a function of doping. One concludes from these curves that
$\bar{g}$ depends only
weakly on the doping and that it is about 0.5 to 1 eV in magnitude. 
If one puts $t$
to zero in the numerator on the right-hand side of Eq.(\ref{coupling})
the ratio is about 0.1 $(eV)^2$. This means that $\bar{g}$ is
determined essentially by $t$ and given approximately by the full band width. 
This result
is a typical strong-coupling result because the effective coupling constant
$\bar{g}$ is determined by the kinetic energy and not by potential
interactions. One also concludes from this that the coupling constant
$g=0.6eV$ used in Ref.\cite{Pines2} for the coupling between holes
and spin excitations is not too large and thus does not cause
the difference in magnitude between the dashed and the dot-dashed
curves in Fig. 4. 
Also shown in Fig. 5 are the lowest eigenvalues of d-wave
symmetry for the total kernel of the O(1) theory (solid line),
the kernel $\Theta_s$ (dot-dashed line, which is identical with the
line representing the numerator on the right-hand side of Eq.(\ref{coupling}))
and the kernel $g^2\chi_P$ (dashed line). The eigenvalues
of the two spin-fluctuation mediated kernels differ by roughly one 
order of magnitude which reflects the ratio of the corresponding curves
in Fig. 4. Using the more realistic value\cite{Balatsky} $\xi \sim 1$ in $\chi_P$
the dashed curve would dramatically move upwards and be near the dotted or
dot-dashed curves. Interesting is also that the eigenvalues of the leading
order of the large-N theory (solid line) is only about a factor 2
larger than those of $\Theta_s$.

Fig. 6 shows curves for the negative imaginary part of the kernel $\Theta$
as a function of frequency. This quantitiy is the analogue of the familiar
function $\alpha^2F(\omega)$ of Eliashberg theory for conventional
superconductivity. The dependences on the momenta $\bf k$ and 
$\bf k'$ have been averaged over the Fermi surface using the 
lowest eigenvalue of d-wave symmetry of the appropriate operator. 
The solid line is the result of the
large-N theory, omitting the instantaneous, attractive term. 
The dot-dashed and the dashed lines correspond to the
spin-fluctuation mediated interactions described by $\Theta_s$ and
$g^2\chi_P$, respectively. 

The d-wave average $<<...>>_3$ over the Fermi line and the momentum
average over the Brillouin zone may yield quite different results
as can be seen by comparing the dashed and dot-dashed curves in
Fig. 1 with the corresponding curves in Fig. 6. Density of states effect
are vital in the average $<<...>>_3$, giving momentum transfer ${\bf q}
\sim {\bf Q}$ a large weight. As a result the dashed and dot-dashed
curves in Fig. 6 resemble more the corresponding curves in Fig. 2
than in Fig. 1. The dashed curve in Fig. 6 exhibits a strong maximum
around 15 meV and a slow decay towards high energies with a typical
decay constant of about 100 meV. Compared to the dashed curve the 
dot-dashed curve in Fig. 6 assumes everywhere only small values, is
rather structureless and has a broad maximum around 0.5 eV. Since the
coupling constants ${\bar g}$ and $g$ are of similar magnitude in the two
cases the difference between the dashed and dot-dashed curves in Fig. 6
is due to the difference in $\chi_P$ and $\chi_\infty$, especially
for momenta ${\bf q} \sim {\bf Q}$.   
If one restricts oneself to energies between zero
and 0.8 eV the integrated spectral weight of the solid
curve is much larger than that of the dot-dashed curve. This implies that the 
next-to-leading contribution in the $1/N$ expansion is substantially smaller 
than the leading one which gives evidence for the usefulness
of the $1/N$ expansion. In the same energy interval the average spectral 
weight of the solid curve is substantially larger than that of the dashed 
curve. However, the solid curve 
becomes strongly negative between 1 and 2 eV. This negative part 
compensates to a large extent the positive
spectral weight between 0 and 0.8 eV if one calculates
the real part of the effective interaction. As a result the retarded O(1)
contribution to $\Theta$ becomes rather small at low frequencies and
the attractive instantaneous part causes mainly the superconductivity,
as discussed in Ref. \cite{Zeyher2}. Such a compensation is absent in
the two spin fluctuation curves because they are always positive
and also very small at high energies.

Using the numerical procedure described in Ref. \cite{Zeyher2} to 
solve the linearized gap equation Eq.(\ref{sup}) we find for T$_c$
the values shown in Fig. 7. The circles describes T$_c$ as a
function of $\delta$ for the O(1) expression for $\Theta$
as discussed in Ref. \cite{Zeyher2}. If only the new term $\Theta_s$
is taken into account the corresponding T$_c$ is given by the 
crosses. These $T_c$'s are at least one order of magnitude 
smaller than the values for $T_c$ using the O(1) expression.
In accordance with that the $T_c$'s for all terms included
in the gap equation (diamonds) are only slightly larger than those
calculated solely from the O(1) contribution.

\section{Imaginary part of the self-energy and relaxation rates}

The particle-particle and the particle-hole effective interactions 
are, except for a possible minus sign in the case of spin fluctuations,
identical if the effective interaction is mediated by bosons with energies
much smaller than the Fermi energy. $T_c$, quasi-particle life times and 
transport relaxation rates are then obtained by taking different Fermi surface
averages of the effective interaction. The particle-particle and the
particle-hole interactions in O(1/N) of the $t-J$ model are, however,
different. Quasi-particle life times and scattering rates are not averages
of the function $\Theta$ of Eq.(\ref{sup}) but of the corresponding
function $\Theta^{ph}$ in the particle-hole channel.
   Following Refs. \cite{Zeyher2,Kulic} the O(1/N) contribution to the
normal-state self-energy $\Sigma'$ of the electronic Green's function $G$, 
defined in Eq.(10) of Ref. \cite{Zeyher2}, is given by

\begin{equation}
\Sigma'(k) = {1 \over N} \sum_{k'} \Theta^{ph}(k,k') g(k+k'),
\label{Sig}
\end{equation}
with
\begin{equation}
\Theta^{ph} = \Theta^{ph}_1 + \Theta^{ph}_2,
\label{sum1}
\end{equation}

\begin{eqnarray}
&\Theta^{ph}_1(k,k') = -\gamma_c(k,k')(t({\bf k}+{\bf k'}) + 
J(-{\bf k'}))
- (t({\bf k}+{\bf k'}) + J(-{\bf k'}))(t({\bf k}) + J({\bf k'}))
\chi_{12}(k')  \nonumber\\
&+ \gamma_c(k,k') \sum\limits^{5}_{r=1} \tilde{E}_r(k) \tilde{\chi}_{2r}(k') - 
\gamma_c(k,k') \sum\limits_{r=1}^{5} \tilde{\chi}_{2r}(k') 
\sum\limits_{s=1}^{5} 
\tilde{\chi}_{rs}(k') \tilde{E}_s(k) \nonumber\\ 
&-\sum\limits_{r,s,t=1}^{5} \tilde{S}_r(k'k) \tilde{\chi}_{rs}(k') 
\tilde{\chi}_{st}(k') \tilde{E}_t(k),
\label{theta_1}
\end{eqnarray}

with 
\begin{equation}
\tilde{S}_r(k',k) = -\sum_{s=1}^5 (1+\tilde{\chi}(k'))^{-1}_{rs}
\tilde{\chi}_{2s}(k')\gamma_c(k,k'),
\label{S}
\end{equation}

\begin{equation}
\Theta^{ph}_2(k.k') = g^{-1}_0(k) Q^{-1} (t({\bf k}+{\bf k'}) +
J(-{\bf k'}))(L({{00\;\;00}\atop{k'}}) + N L({{12\;\;21}\atop{k'}})).
\label{theta_2}
\end{equation}
L is the Fourier transform of the equilibrium Green's function of two 
bosonic operators $X(1)$, $X(2)$, except for a minus sign,
\begin{equation}
L(1,2) = \langle  T
(X(1) - \langle X(1) \rangle )
(X(2) - \langle X(2) \rangle ) \rangle.
\label {Green}
\end{equation}
$g_0(k)$ is the free quasi-particle Green's function $(i\omega_n -
\epsilon({\bf k}))^{-1}$ and $Q$ the spectral weight 
$\langle X^{00}({\bar 1}) \rangle + \langle X^{11}({\bar 1}) \rangle $.
At large N the second term in $Q$ could be dropped because it contributes
only higher orders in 1/N than those considered here. For the transition
from large N to the physical value $N=2$ it is advisable, however, to keep
also higher orders in 1/N in the inverse spectral weight in order to
improve the convergence of the 1/N expansion at small doping. The
remaining symbols in Eqs.(\ref{theta_1}), (\ref{S}), and (\ref{theta_2}) have 
been defined in Eqs.(41-48) of Ref. \cite{Zeyher2}. 

The self-energy $\Sigma'$ is associated with
the physical Green's function $G$. Taking only $\Theta_1^{ph}$ in 
Eq.(\ref{sum1}) into account Eq.(\ref{Sig}) defines a different self-energy 
$\Sigma$ related to the quasi-particle Green's function $g$ defined
in Eq.(15) of Ref. \cite{Zeyher2}. If $k$ is near the Fermi surface
(i.e., $\epsilon({\bf k})$ near the Fermi energy and the frequency $i\omega_n$
small) $\Theta^{ph}_2$ of Eq.(\ref{theta_2}) is small due to the factor 
$g_0^{-1}(k)$. This means that the two kind of self-energies are
essentially equal near the Fermi surface. However they differ
considerably far away fom the Fermi energy: The imaginary part of
$\Sigma'$ is there always negative definite as it should be in contrast 
to that of $\Sigma$ which assumes there in general both signs.

Fig. 8 shows various contributions to the negative imaginary part of
$\Sigma'$ as a function of energy
for $t=0.4 eV$, $J/t=0.3$, $\delta = 0.17$, calculated at very
low temperatures. 
The momentum in the self-energy has always been averaged over the Fermi 
surface. The dotted line
describes the pure spin-flip contribution, i.e., the second term in 
Eq.(\ref{theta_1}) and the second L-contribution in Eq.(\ref{theta_2}).
The dot-dashed line in Fig. 8 contains the charge contribution to the
self-energy, i.e., the first term in Eq.(\ref{theta_1}) and the first
L-contribution in Eq.(\ref{theta_2}). The solid line in Fig. 8 contains the
contribution of all terms in $\Theta^{ph}$, i.e., the full $O(1)$ contribution
of $\Sigma'$. First we note that the sum
of the dotted and dot-dashed lines in Fig. 8 is not much different from
the solid line. This means that the terms 3-5 in Eq.(\ref{theta_1})
describing mixed charge and spin fluctuations yield only a small contribution
to $\Sigma'$. The imaginary part of $\Sigma'$ shows a double peak both
for negative and positive frequencies. The peaks at lower frequencies 
are due to spin, the peaks at higher energies due to 
collective charge fluctuations. This is in agreement with the properties
of the spin and charge excitation spectra of the $t-J$ model \cite{Gehlhoff}.
For relaxation rates and transport only the low-frequency part
of the imaginary part of $\Sigma'$ is relevant. Fig. 8 suggests that
this part is mainly determined by spin and only to a lesser degree
by charge fluctuations.

Theories for high-T$_c$ superconductors which are based on a retarded,
boson-mediated pairing mechanism usually encounter the problem to
reconcile the large values for T$_c$ with the rather modest value for
the resistivity. Using the leading order of the 1/N expansion it was
shown\cite{Zeyher2} that the obtained values for $T_c$ are comparable with 
those found in high-$T_c$ superconductors. The question immediately arises
whether the same theory (and also theories based on phonon- or
spin fluctuation-mediated interactions) are in agreement with the observed
low electronic scattering rates in these systems. 
For this aim we study the quantities $1/\tau$ and $1/\tau_{tr}$
which describe the relaxation of quasi-particles 
and the momentum relaxation near the Fermi surface, the latter being intimately
related to transport. Using our previously defined effective
particle-hole interaction $\Theta^{ph}$ one obtains \cite{Allen}

\begin{equation}
1/\tau = {2 \over N} \int_{-\infty}^{\infty} d \epsilon N(\epsilon)
<<Im\Theta^{ph}({\bf k},{\bf k'}-{\bf k},\epsilon)>>_0
(n_b(\epsilon)+n_F(\epsilon)),
\label{tau1}
\end{equation}
\begin{equation}
1/\tau_{tr} = {2 \over N} \int_{-\infty}^{\infty} d \epsilon N(\epsilon)
{{<<Im\Theta^{ph}({\bf k},{\bf k'}-{\bf k},
\epsilon)({\bf v}({\bf k})-{\bf v}({\bf k'}))^2>>_0}\over
{<< ({\bf v}({\bf k})-{\bf v}({\bf k'}))^2>>_0}}
(n_b(\epsilon)+n_F(\epsilon)),
\label{tau2}
\end{equation}
$<<...>>_0$ denotes an average over the Fermi line with a constant 
eigenvector, i.e., the weight in the line integral around the Fermi line
contains only the density of states. $Im$ stands for the imaginary part, 
${\bf v}({\bf k})= {\partial \epsilon({\bf k})/{\partial {\bf k}}}$ 
is the velocity of the electron at the point ${\bf k}$. $N(\epsilon)$,
$n_B(\epsilon)$ and $n_F(\epsilon)$ denote the density of states, the Bose 
and the Fermi distribution function, respectively.
In the high temperature limit $1/\tau$ and $1/\tau_{tr}$ approach the limits
$2\pi T\lambda$ and $2\pi T\lambda_{tr}$, respectively, where $\lambda$ and
$\lambda_{tr}$ are the corresponding dimensionless coupling strengths.
In order to achieve the desired
high values for $T_c$ with a spin-fluctuation or phonon mechanism 
$\lambda$ has to be roughly between two and three
(2.6 in Ref.\cite{Carbotte}, 3 in Ref.\cite{Collins}, 2.28 
using $\chi_P$ and the values of Table II, second-last column, of Ref.
\cite{Pines3}, 2.9 in Ref. \cite{Zeyher4}). On the other hand, 
infrared and resistivity data in many cuprates near optimal
doping lead to a much lower experimental value for 
$\lambda_{tr}$ of about $0.3\pm 0.1$, see Table II in Ref.\cite{Tanner}.

The dashed curves in Figs. 9 and 10 
show $1/\tau$ and $1/\tau_{tr}$, respectively,
as a function of temperature for the kernel $g^2\chi_P$. The curves are
quasilinear above 150 or 200 K corresponding to dimensionless
coupling constants $\lambda_P$ and $\lambda_{tr,P}$ of about 2-3.
Such a large value for $\lambda_P$ is needed in order to
reproduce the large value for $T_c$. The corresponding transport
value $\lambda_{tr,P}$, however, is even larger by about 20 per cent. 
The reason
for this is the following: If $\Theta$ is momentum independent $\lambda$
and $\lambda_{tr}$ are equal. Considering now kernels $\Theta$ which 
depend only on the difference of their momenta, i.e., ${\bf k}-{\bf k'}$.
If $\Theta$ shows mainly forward scattering,
i.e., if $Im \Theta({\bf k}-{\bf k'})$ is large for small momentum and small
at large momentum transfers $\lambda_{tr}$ is smaller than $\lambda$.
On the other hand if scattering occurs mainly with large momentum transfers
$\lambda$ is smaller than $\lambda_{tr}$. Spin-fluctuations are 
strongest in the region around ${\bf Q}$ in momentum space, i.e.,
at large momentum transfers. As a result $\lambda_{tr,P} > \lambda_P$ in
agreement with the slopes in Figs. 9 and 10. We expect that 
$\lambda_{tr,P} > \lambda_P$
and $\lambda \sim 2-3$ are generic values for spin fluctuation models
for high-T$_c$ superconductivity. Thus 
there is a severe discrepancy of about one order of magnitude between
theoretical and experimental values for $\lambda_{tr,P}$.

The long-dashed curves in Figs. 9 and 10 describe relaxation rates due to 
phonons. Following our previous 
treatment \cite{Greco2} we used a Holstein model with a phonon frequency 
$\omega_0= 48 meV$ and a dimensionless coupling strength $\lambda_{Ph} = 0.75$
near optimal doping. 
As discussed in Refs.\cite{Zeyher3,Greco2} vertex corrections must be included
even in the leading order of the $1/N$ expansion. The resulting vertex,
taken at $\omega=0$, depends strongly on momentum in the region of the 
considered doping values. This implies that the constant, bare electron-phonon
coupling becomes effectively strongly momentum dependent, being large at 
small and
small at large momentum transfers. As a result the phonon contribution 
to $\lambda_{tr,Ph}$, given essentially by the slope of the long-dashed line 
in Fig. 10, becomes
much smaller\cite{Zeyher3} than $\lambda_{Ph}=0.75$, namely, 
$\lambda_{tr,Ph} \sim 0.11$. More generally, such a strong reduction in 
$\lambda_{tr}$ relative to $\lambda$, should occur for all charge 
but not for spin interactions. This may explain why phonons and impurities
play usually no role in transport data though their bare couplings
to electrons may not be small\cite{Andersen}. 

The relation between $T_c$ and relaxation rates is quite different in the case of the $1/N$ expansion. The 
instantaneous part of the kernel is large, for instance, its Fermi surface
average in the d-wave channel is -0.48 at $\delta = 0.17$. This term alone
would yield a $T_c$ of about the same magnitude as the whole O(1)
contribution for $\Theta$. On the other hand, this instantaneous term
does not contribute to the momentum relaxation , i.e., it does not enter
transport. The solid lines in Figs. 9 and 10 represent the relaxation rates
due to $\Theta^{ph}$ as given by Eqs.(\ref{sum1}-\ref{theta_2}). The
curves are again quasilinear at not too low temperatures. Their absolute
values are rather small and correspond to dimensionless coupling
constants of $\lambda \sim 0.25$ and $\lambda_{tr} \sim 0.22$. 
This means that charge and spin fluctuations do not scatter electrons
very efficiently
near the Fermi surface. Finally, the long-dashed curves in Figs. 9 and 10
are the relaxation rates due to the spin fluctuation kernel $\Theta_s$
of the 1/N expansion. Comparing these curves to the dashed curves in
Figs. 9 and 10 one recognizes that the leading spin fluctuation contribution,
which is of O(1/N) in $\Theta$, yields only small scattering rates,
compared to those in spin fluctuation models for high-$T_c$
compounds.  The total transport
coupling constant is the sum of its electronic and phononic parts and
about 0.5 in our model. This value is near the rather universal experimental 
value of $0.3\pm 0.1$\cite{Tanner}.

\section{Conclusions}

Using a $1/N$ expansion for the $t-J$ model in terms of X operators the 
O(1/N) contributions to the anomalous self-energy just below $T_c$ and
the self-energy in the normal state have been considered and numerically
evaluated for typical values of the model parameters. Our treatment is
restricted to the optimally doped and the overdoped region defined by the
absence of additional
instabilities, e.g., towards a flux state.
Unlike to the case of phonons or spin fluctuations 
the resulting effective interactions in the particle-particle and 
particle-hole channels turn out to be different from each other. We also 
show how
scattering rates relevant to transport can be obtained form the expression
for the self-energy. In order to examine the convergence of the $1/N$ 
expansion and to take into account antiferromagnetic effects 
we also have included graphs of $O(1/N^2)$ which
describe spin-fluctuation mediated contributions to the effective
interaction between electrons.  Though these terms are only a subset of all
$1/N^2$ contributions it seems not unreasonable to assume that they
represent the leading corrections to the O(1/N) terms if, as is often
believed, spin fluctuations play an important role in the $t-J$ model in the 
considered doping region.

Our main conclusion is that the leading O(1/N$^2$) contribution describing
spin fluctuations gives only a small correction to the effective 
interaction and $T_c$ obtained already in the leading order O(1/N). 
As discussed in Ref.\cite{Zeyher2} the effective interaction in O(1/N)
contains no RPA-like spin fluctuations. Instead, this effective interaction
and thus also d-wave pairing is
caused by the instantaneous contribution of the Heisenberg
term and by spin flip processes induced by the X operator algebra and
thus by the constraint. The reason for the smallness of the
O(1/N$^2$) contribution 
is that the momentum averaged magnetic susceptibility at large N's, 
$\chi_\infty$, is, as a function of frequency, rather structureless and 
extends over the whole bandwidth. Though the effective coupling constant
between electrons and spin excitations is set by $t$ ( and not $J$) the
resulting contribution to the imaginary part of the effective interaction
(the analogue of $\alpha^2F$ of the phonon case) is small compared to
the spin flip (and charge) contribution obtained in O(1/N). In contrast to
that the spin susceptibilities used in spin fluctuations theories such as
$\chi_P$ of Refs.\cite{Pines1,Pines2,Pines3} 
is large and strongly peaked at low frequencies. In other words, most of
the spectral weight is in this case accumulated at small frequencies 
of the order of a fraction of $J$ whereas it is rather constant and spread 
over 
an energy scale of $t$ in the case of $\chi_\infty$.     
Available magnetic susceptibility spectra from inelastic neutron scattering
in $YBa_2Cu_3O_{7-x}$ refer mostly to the strongly underdoped
region and resemble $\chi_P$ more than $\chi_\infty$. 
In optimally doped samples of $YBa_2Cu_3O_{7-x}$ spin scattering is too weak 
to be detected presently. The obtained experimental upper bound for $\chi''$, 
however,
suggests in view of the sum rule that $\chi''$ is rather
structureless and spreads over an energy scale set by $t$ similar as
$\chi''_\infty$ does. There is another fundamental difference between
calculations based on 1/N expansions and spin fluctuation models:
The effective interaction of O(1/N) in the d-wave channel contains
strongly pair-breaking charge excitations at high energies of the order $t$
which cancel the contribution of pair-forming low-energy spin flip 
excitations to a large extent. These pair-breaking terms are entirely
neglected in the spin fluctuation models but play an important role
for quantitative calculations of $T_c$ in the 1/N expansion. 

For the same reasons as above we find that the leading spin fluctuation 
terms of O(1/N$^2$) modify only little the scattering rates $1/ \tau$ 
and $1/\tau_{tr}$ obtained already in O(1/N). If spin fluctuation terms would
dominate and cause the large values of $T_c$ the corresponding 
scattering rate $1/\tau_{tr}$
would exceed the experimental ones by far. Responsible for this is also 
the
fact that the spin susceptibility is large for large momentum transfers
which makes $1/\tau_{tr}$ even larger than $1/\tau$. In the 1/N expansion
large values for $T_c$ are compatible with small values for $1/\tau_{tr}$:
Spin fluctuation terms are small and $T_c$ is to a large extent determined
by the instantaneous part of the interaction which does not contribute to  
relaxation rates. The case of phonons or, more generally, of charge-charge
interactions is somewhat different from spin interactions. Here the 
charge vertex due to correlation effects has to be included already in 
leading order of the 1/N expansion. As a result collective effects 
dominate and charge interactions
are confined to small momenta causing  $1/\tau_{tr} << 1/\tau$
near optimal doping. This may explain why phonons and impurities
play only a minor role in the transport data of optimally doped high-$T_c$
compounds.

{\bf Acknowledgement}: The authors are grateful to Secyt and the International
Bureau of the Federal Ministry for Education, Science, Research and 
Technology of Germany for financial support (Scientific-technological 
cooperation between Argentina and Germany, Project No. ARG AD 3P).
The first and second author thank the MPI-FKF, Stuttgart, Germany, and
the Departamento de F\'{\i}sica, Facultad de Ciencias Exactas e 
Ingenier\'{\i}a, Rosario, Argentina, respectively, for hospitality. 
The authors also thank P. Horsch for a critical reading of the manuscript.

\vspace{2cm}

%\begin{references}

\newpage

\begin{centerline}
{\large {FIGURE CAPTIONS}}
\end{centerline}
\vspace{1cm}

\noindent
Fig.1:
Momentum integrated magnetic susceptibility $\chi''(\omega)$ in units of 
$\mu^2_B/eV$. Dot-dashed line: $t-J$ model for $N \rightarrow \infty$;
dashed line: $\chi_P''(\omega)$ of Ref.\cite{Pines3}; circles: exp.
points for $YBa_2Cu_3O_{6.6}$ at $T=80K$\cite{Hayden}.\\

\noindent
Fig.2:
Magnetic susceptibility $\chi''({\bf Q},\omega)$ for ${\bf Q}=(\pi,\pi)$
in units of $\mu^2_B/eV$. Dot-dashed line: $t-J$ model for $N \rightarrow 
\infty$ multiplied by 10;
dashed line: $\chi_P''({\bf Q},\omega)$ of Ref.\cite{Pines3}; circles: exp.
points for $YBa_2Cu_3O_{6.7}$ at $T=200K$\cite{Fong}.\\

\noindent
Fig.3:
Absolute value of the real part of the susceptibility at $\omega=0$
versus distance. Circles
and diamonds correspond to $\chi_\infty$ and $\chi_P$, respectively,
at neighboring sites of the square lattice, the dashed and dot-dashed lines to
$K_0(x/\xi)$ with $\xi=2.3$ and $\xi=1.0$, respectively. \\

\noindent
Fig.4:
Dependence of the static kernel $\Theta({\bf k},{\bf k'},0)$ on the second
momentum ${\bf k'}$ along the Fermi line for a fixed first momentum
${\bf k}=(2.465,0.309)$; solid line: O(1) contrib.;
dot-dashed line: O(1/N) contrib.; dashed line: $g^2\chi_P$ divided
by 8.\\ 

\noindent
Fig.5:
Lowest eigenvalue of static kernels in $\Gamma_3$ (d-wave) symmetry; 
solid, dashed, dot-dashed, and dotted lines correspond to O(1) of $\Theta$,
$g^2\chi_P$, the numerator, and denominator of Eq.(18), respectively.
The dot-dashed curve corresponds at the same time to $\Theta_s$.\\

\noindent
Fig.6:
Negative imaginary part of the d-wave projected kernel 
$\Theta_3(\omega+i\eta)$ as a function of the frequency; solid, 
dot-dashed, and dashed lines correspond to the contributions of O(1), 
O(1/N), and $g^2\chi_P$ to $\Theta_3$, respectively. \\   

\noindent
Fig.7:
Transition temperature $T_c$ in Kelvin as a function of doping; circles,
crosses, and diamonds correspond to the contributions of O(1), O(1/N),
and O(1)+O(1/N), respectively.\\

\noindent
Fig.8:
Negative imaginary part of the O(1/N) electronic self-energy as a function
of energy with the momentum averaged over the Fermi surface. Dot-dashed,
dashed, and solid lines describe the charge, spin and total contributions,
respectively.\\

\noindent
Fig.9:
Quasi-particle relaxation rate $1/\tau$ as a function of temperature.
Dashed, long-dashed, solid, and dot-dashed lines correspond to $\chi_P$,
an electron-phonon model, and the O(1/N) and O(1/N$^2$) contributions
to the self-energy of the $t-J$ model, respectively.\\

\noindent
Fig.10:
Transport relaxation rate $1/\tau_{tr}$ as a function of temperature.
Dashed, long-dashed, solid, and dot-dashed lines have been calculated
from $\chi_P$, an electron-phonon model, and the O(1/N) and O(1/N$^2$) 
contributions to the self-energy of the $t-J$ model, respectively.\\

\newpage

\begin{figure}[h]
\centerline{\psfig{figure=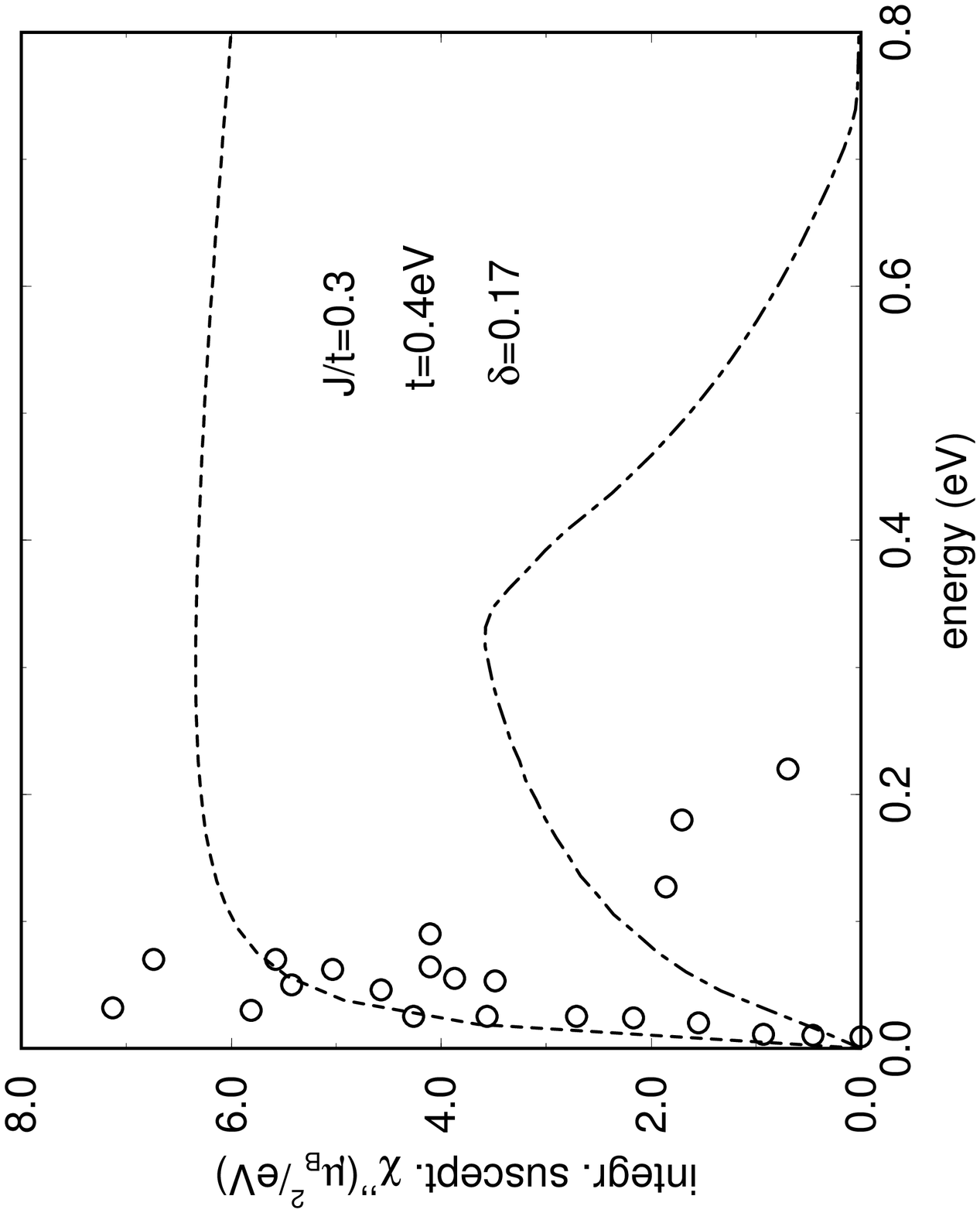,width=18cm}}
\caption{}
\end{figure}
\newpage

\begin{figure}[v]
\centerline{\psfig{figure=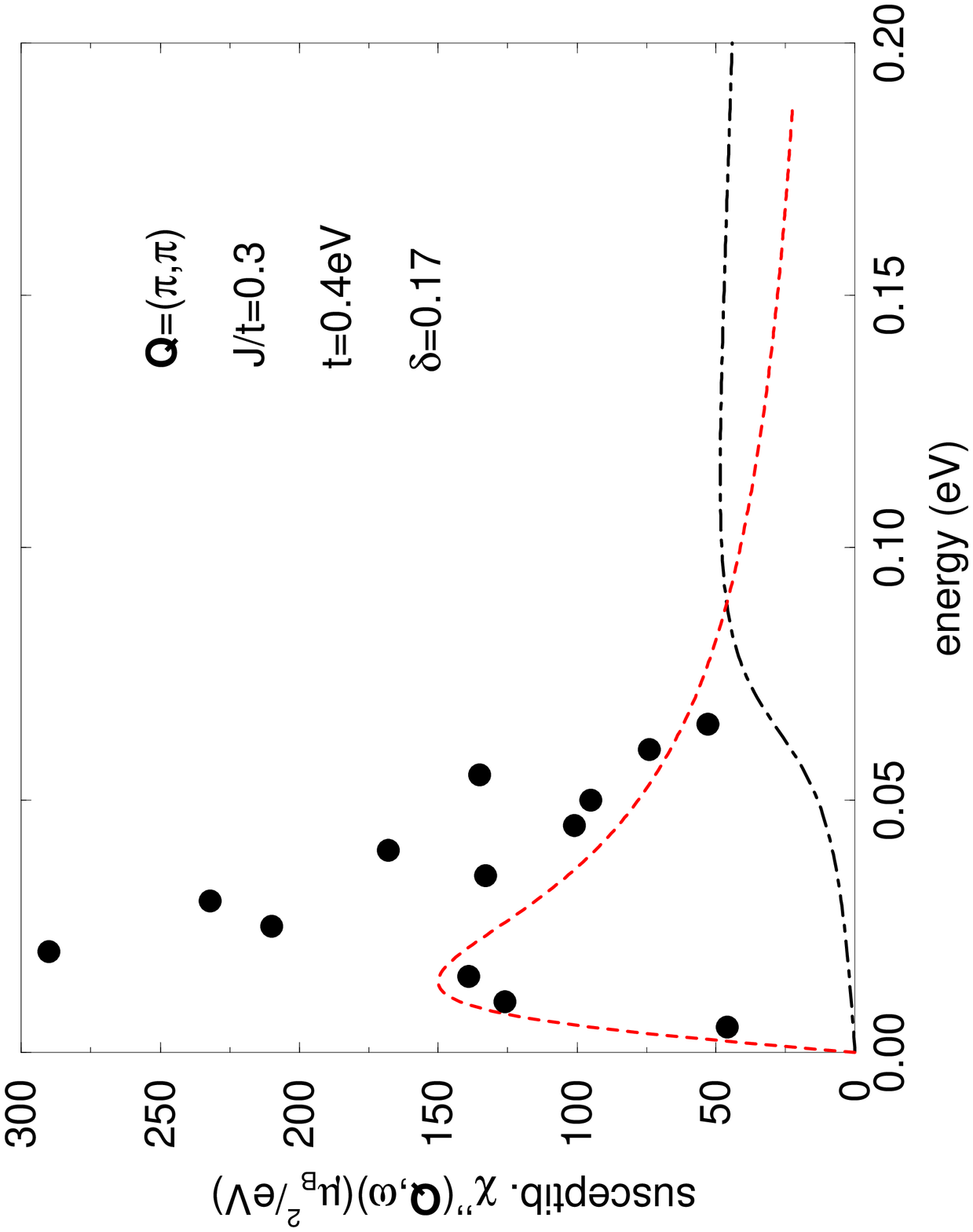,width=18cm}}
\caption
{}
\end{figure}
\newpage

\begin{figure}
\centerline{\psfig{figure=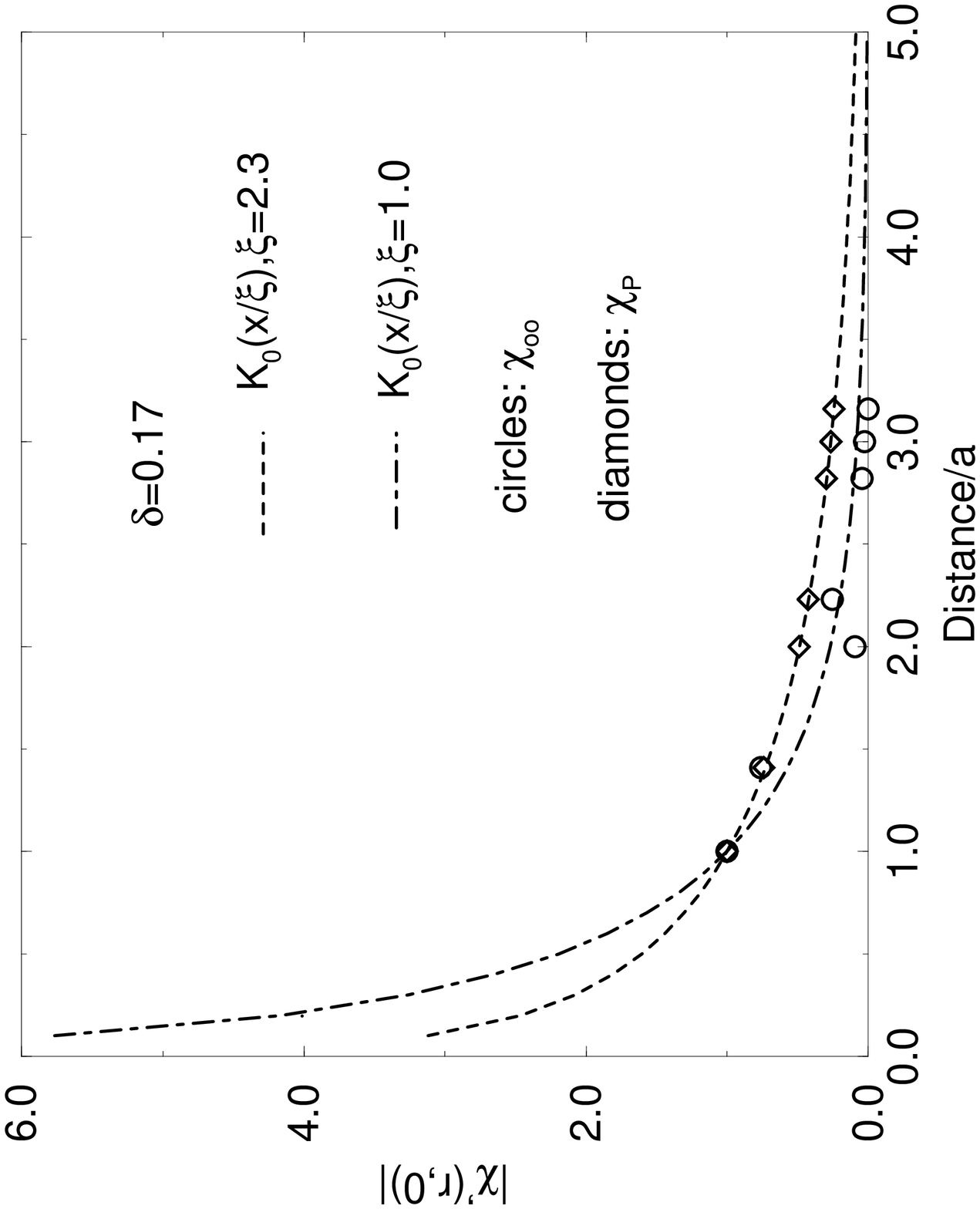,width=18cm}}
\caption
{}
\end{figure}
\newpage

\begin{figure}
\centerline{\psfig{figure=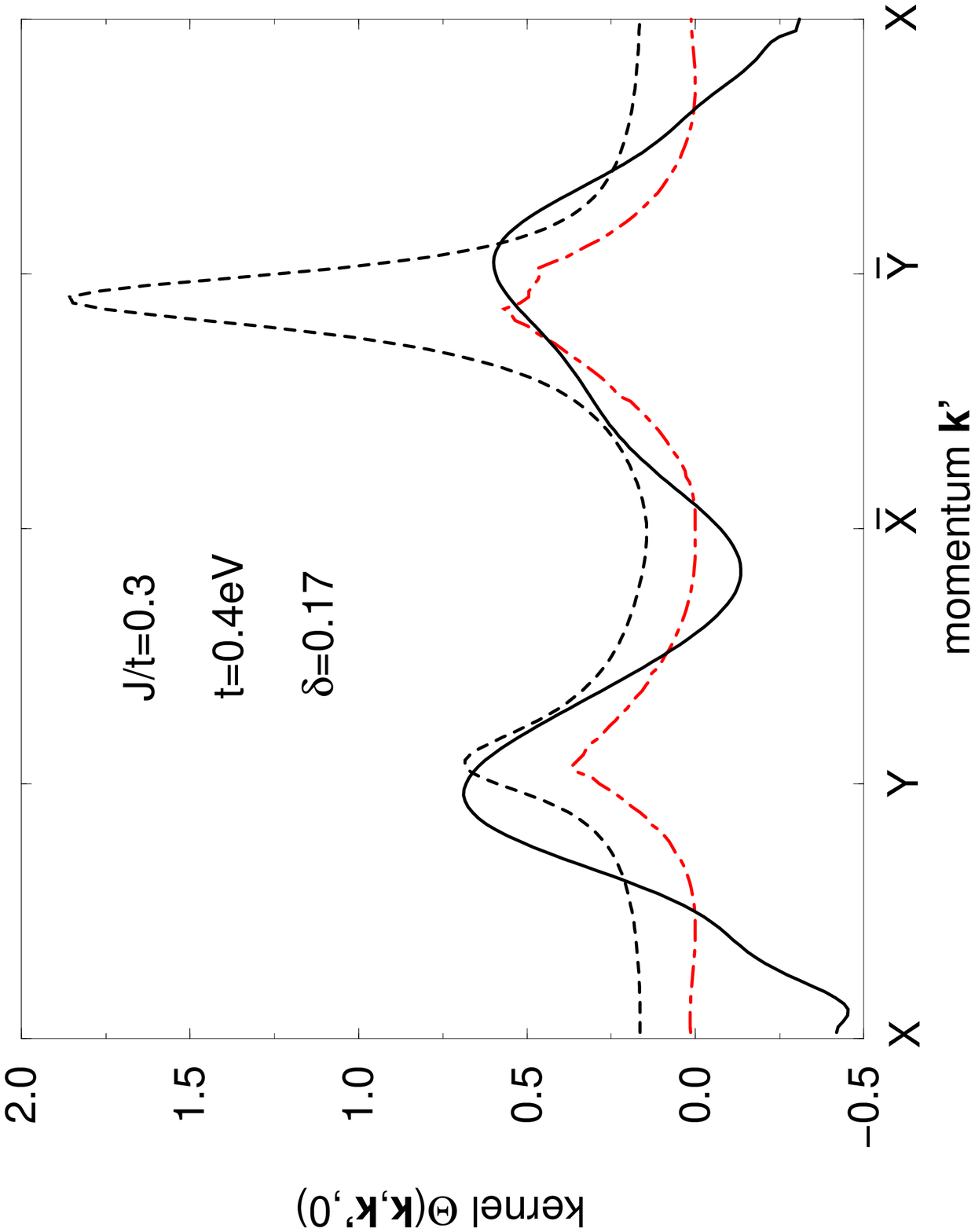,width=18cm}}
\caption
{}
\end{figure}
\newpage

\begin{figure}
\centerline{\psfig{figure=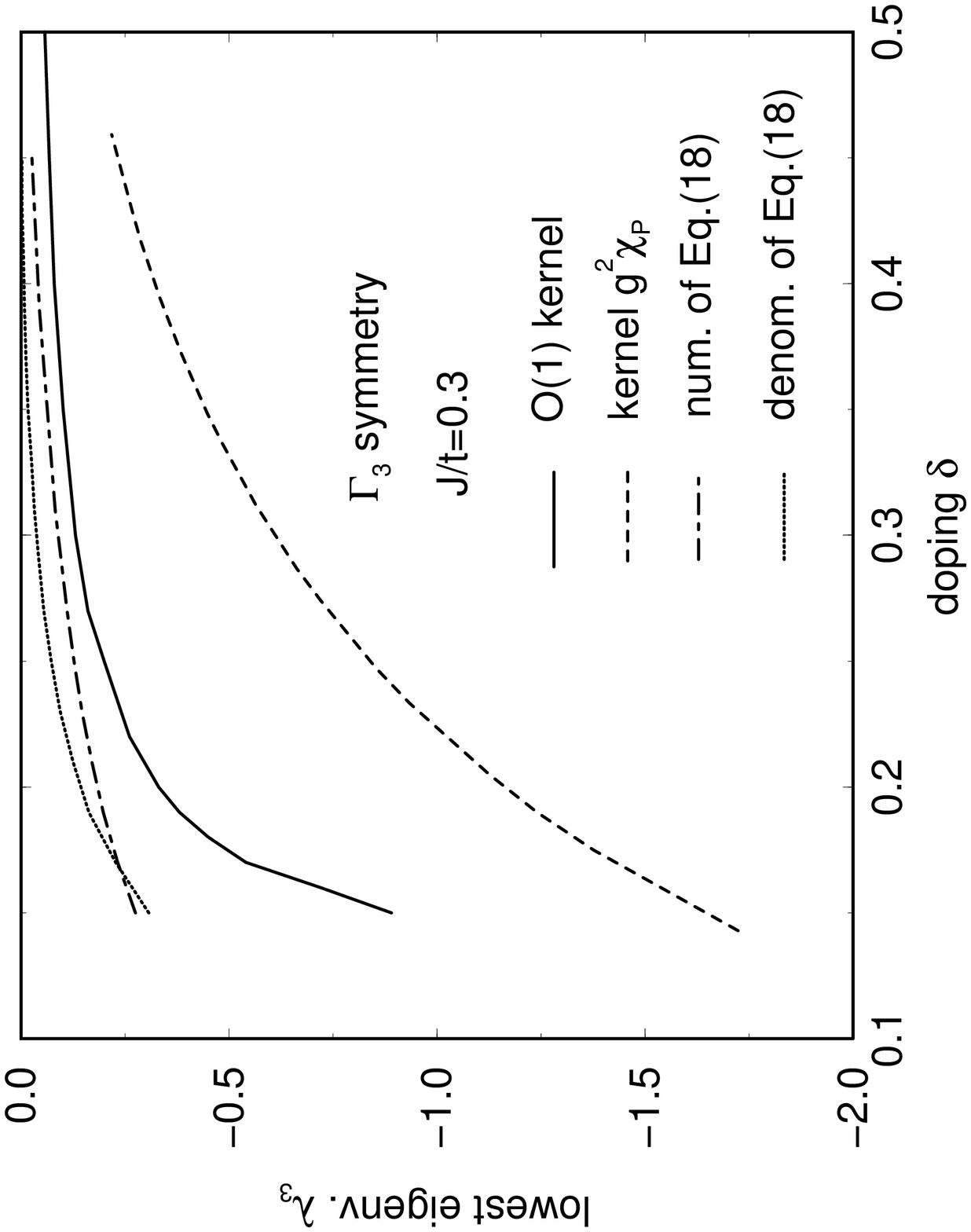,width=18cm}}
\caption
{}
\end{figure}
\newpage

\begin{figure}
\centerline{\psfig{figure=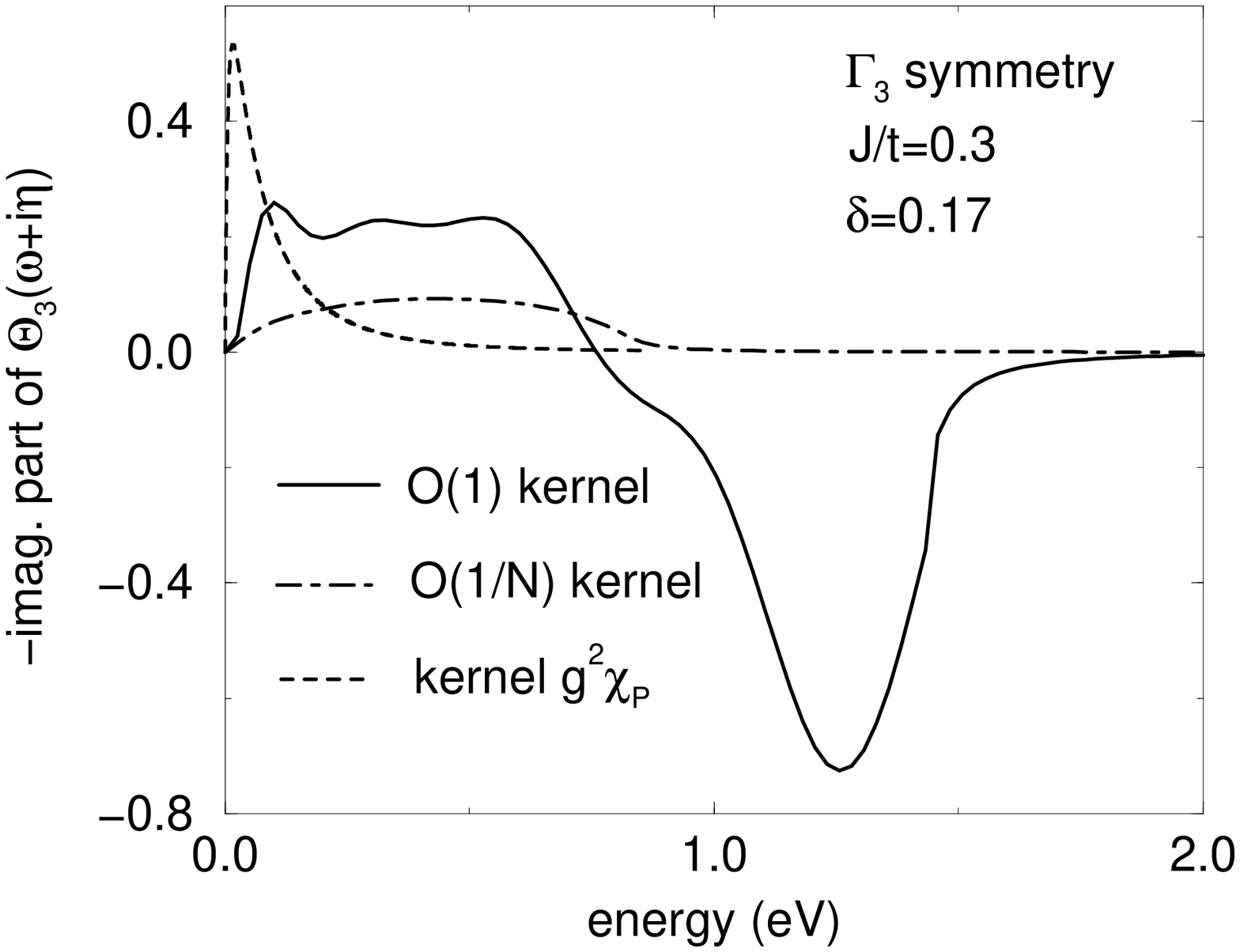,width=18cm}}
\caption
{}
\end{figure}
\newpage

\begin{figure}
\centerline{\psfig{figure=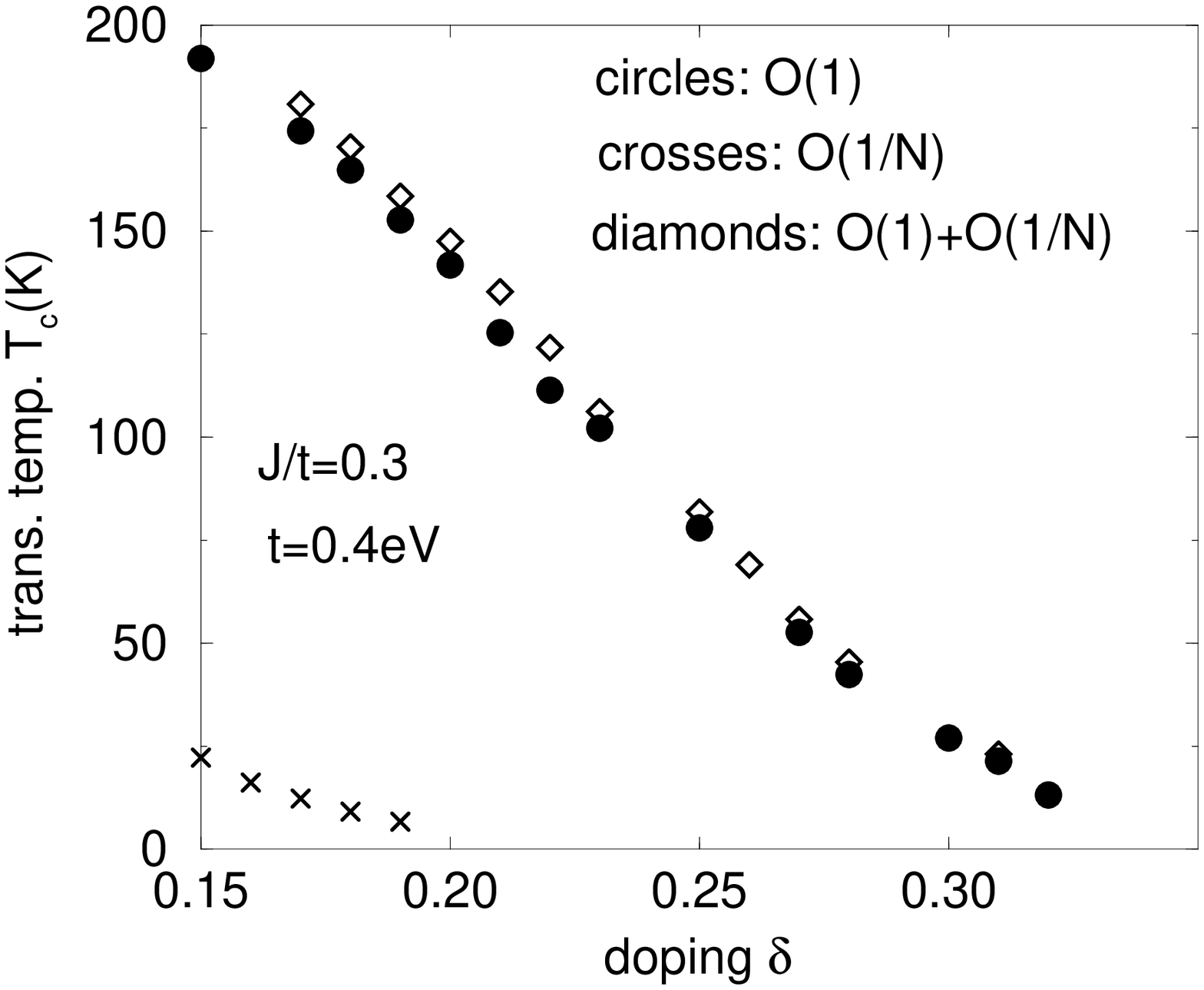,width=18cm}}
\caption
{}
\end{figure}
\newpage

\begin{figure}
\centerline{\psfig{figure=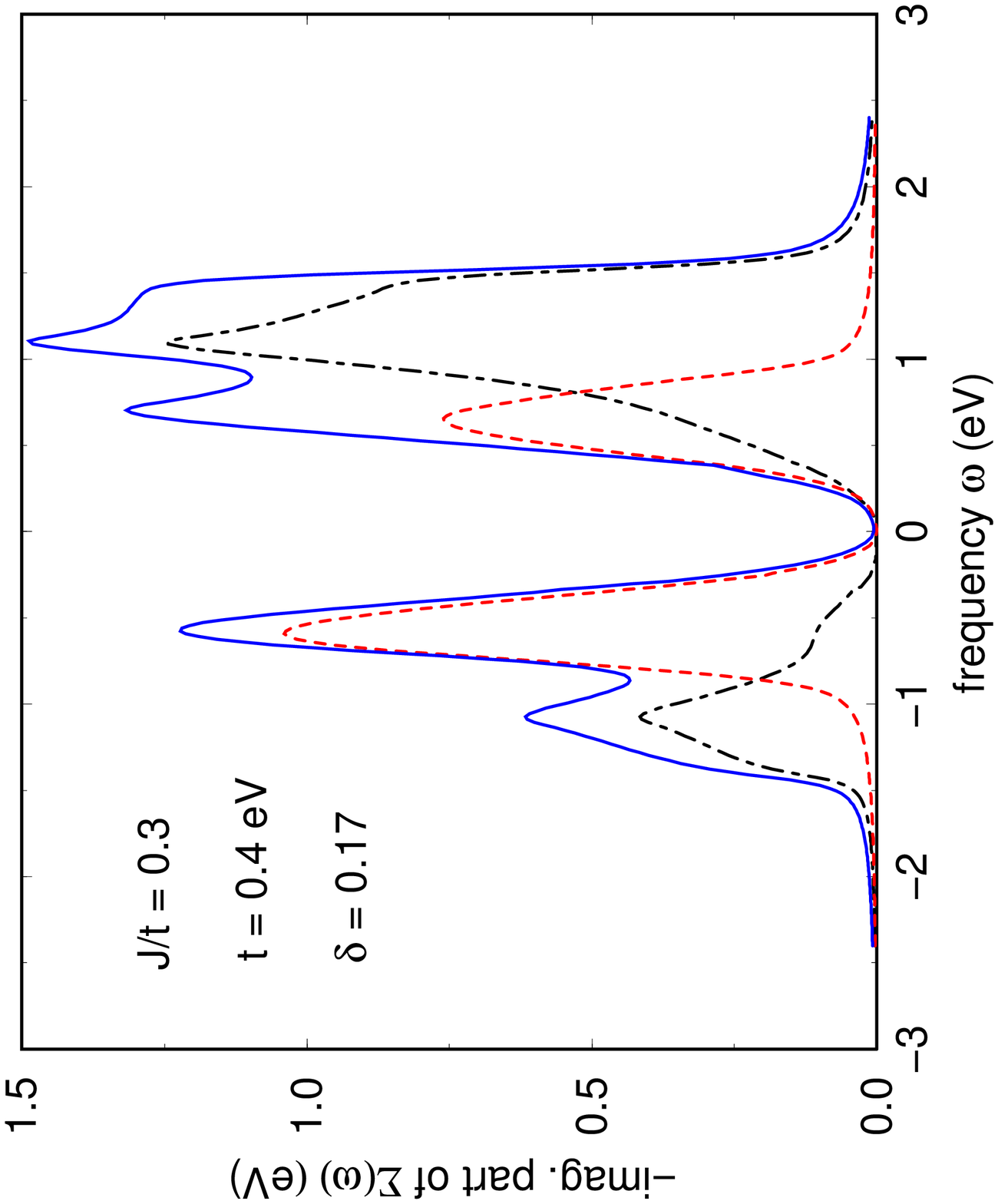,width=18cm}}
\caption
{}
\end{figure}
\newpage

\begin{figure}
\centerline{\psfig{figure=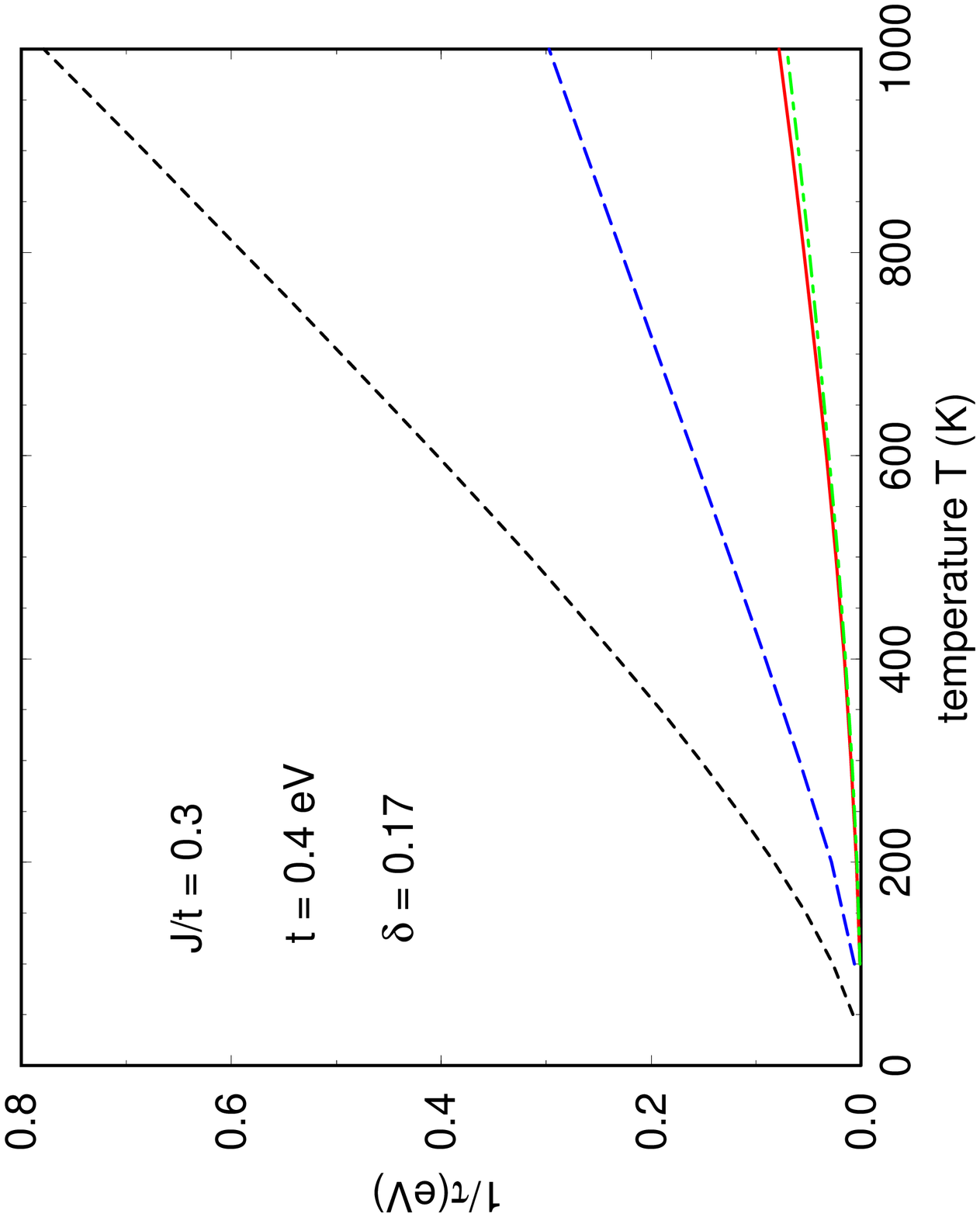,width=18cm}}
\caption
{}
\end{figure}
\newpage

\begin{figure}
\centerline{\psfig{figure=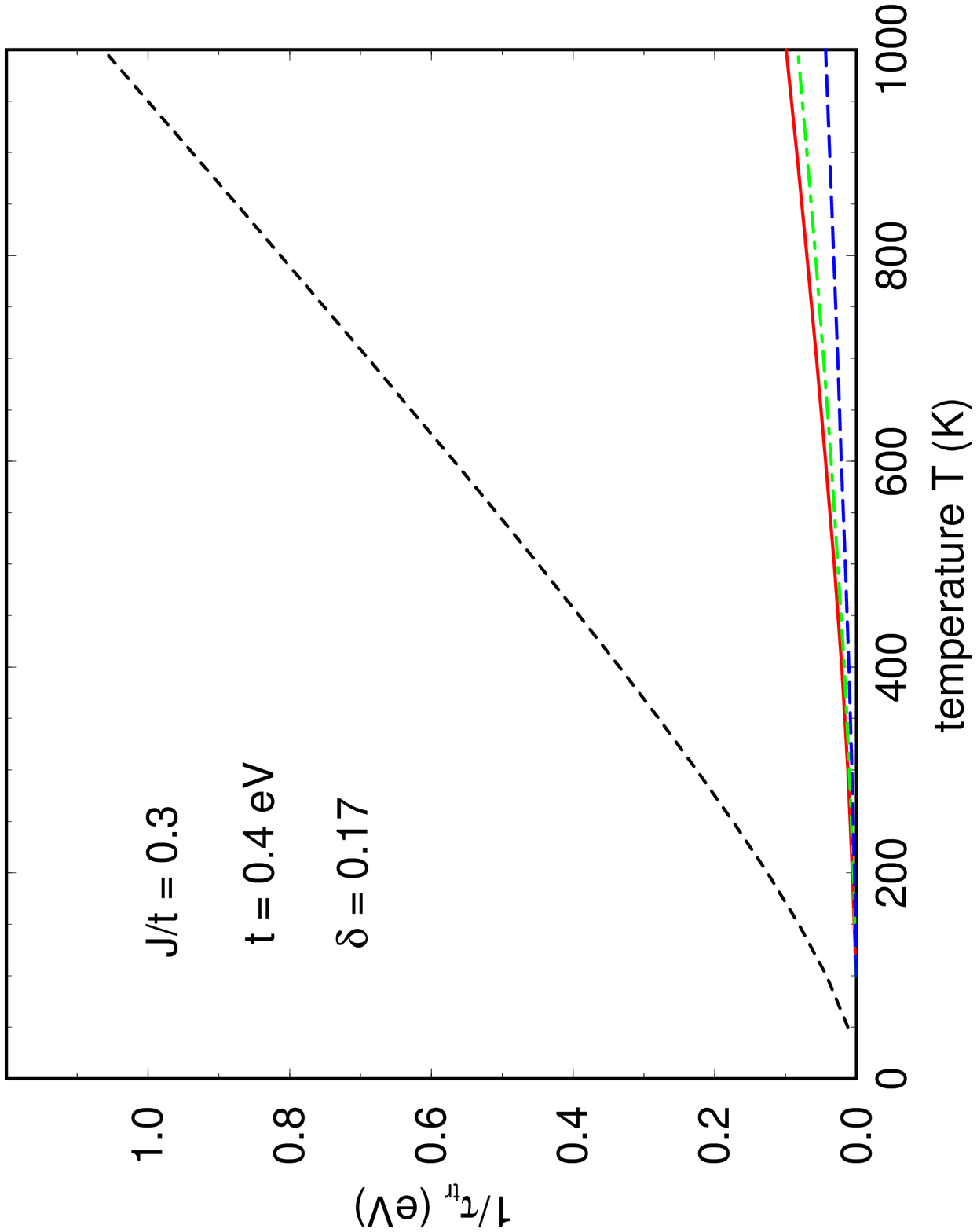,width=18cm}}
\caption
{}
\end{figure}
\newpage


\begin{thebibliography}{99}

\bibitem{Bickers} N.E. Bickers, D.J. Scalapino, and S.R. White, Phys. Rev.
Lett. {\bf 74}, 961 (1989)

\bibitem{Grabowski} S. Grabowski et al., Europhys. Lett. {\bf 34}, 219 (1996)

\bibitem{Greco1} A. Greco and R. Zeyher, Europhys. Lett. {\bf 35}, 115 (1996)

\bibitem{Zeyher1} R. Zeyher and A. Greco, Z. Phys. B {\bf 104}, 737 (1997)

\bibitem{Zeyher2} R. Zeyher and A. Greco, Eur. Phys. J B {\bf 6}, 473
(1998)

\bibitem{Pines1} P. Monthoux, A. Balatsky, and D. Pines, 
Phys. Rev. B {\bf 46}, 14803 (1992)

\bibitem{Pines2} P. Monthoux and D. Pines, Phys. Rev. B {\bf 47}, 
6069 (1993)

\bibitem{Pines3} P. Monthoux and D. Pines, Phys. Rev. B {\bf 49}, 
4261 (1994)
 
\bibitem{Carbotte} J.P. Carbotte, E. Schachinger, and D.N. Basov, 
Nature {\bf 401}, 354 (1999)

\bibitem{Morse} D.C. Morse and T. Lubensky, Phys. Rev. B{\bf 42},7794 (1990)

\bibitem{Cappelluti1} E. Cappelluti and R. Zeyher, Phys. Rev. B {\bf 59},
6475 (1999)

\bibitem{Kulic} M.L. Kuli\'c and R. Zeyher, Mod. Phys. Lett. B {\bf 11},
333 (1997)

\bibitem{Cappelluti2} E. Cappelluti and R. Zeyher, Int. J. Mod. Phys.
B{\bf 13}, 2607 (1999) 

\bibitem{Hayden} S.M. Hayden, G. Aeppli, P. Dai, H.A. Mook, T.G. Perring, 
S.-W. Cheong, Z. Fisk, F. Dogan, and T.E. Mason, Physica B {241-243},
765 (1998) 

\bibitem{Bourges1} Ph. Bourges, in {\it The Gap Symmetry and Fluctuations
in High Temperature Superconductors}, edited by J. Bok, G. Deutscher,
D. Pavuna, and S.A. Wolf, Plenum, New York (1998), p. 349

\bibitem{Bourges2} Ph. Bourges, cond-mat/9902067 

\bibitem{Fong} H.F. Fong, P. Bourges, Y. Sidis, L.P. Regnault, J. Bossy, 
A. Ivanov, D.L. Milius, I.A. Aksay, and B. Keimer, Phys. Rev. B {\bf 61},
14773 (2000)

\bibitem{Aeppli} G. Aeppli, S.M. Hayden, P. Dai, H.A. Mook, R.D. Hunt,
T.G. Perring, and F. Dogan, phys. stat. sol. (b) {\bf 215}, 519 (1999)

\bibitem{Sidis} Y. Sidis et al., Phys. Rev. Lett. {\bf 84}, 5900 (2000)

\bibitem{Balatsky} A.V. Balatsky and P. Bourges, Phys. Rev. Lett. {\bf 82},
5337 (1999)

\bibitem{Gorny} K.R. Gorny, O.M. Vyaselev, S. Yu, C.H. Pennington, W.L. Hults,
and J.L. Smith, Phys. Rev. Lett. {\bf 81}, 2340 (1998)

\bibitem{Berthier} T. Auler, M. Horvatic, J.A. Gillet, C. Berthier, 
Y. Berthier, P. Carretta, Y. Kitaoka, P. S\'egransan, J.Y. Henry,
Physica C 313, 255 (1999)

\bibitem{Levin1} Q. Si, Y. Zha, K. Levin, and J.P. Lu, 
Phys. Rev. B {\bf 47}, 9055 (1993)

\bibitem{Bulut} N. Bulut, D.J. Scalapino, and S.R. White, 
Phys. Rev. B {\bf 47}, 2742 (1993)

\bibitem{Gehlhoff} L. Gehlhoff and R. Zeyher, Phys. Rev. B {\bf 52},
4635 (1995)

\bibitem{Allen} P.B. Allen and B. Mitrovi$\acute{c}$, Sol. State Physics
{\bf 37}, 1 (1982)

\bibitem{Zeyher4} R. Zeyher and G. Zwicknagl, Z. Phys. B - Condensed
Matter {\bf 78}, 175 (1990)

\bibitem{Tanner} D.B. Tanner and T. Timusk, in ``Physical Properties
of High-T$_c$ Temperature Superconductors III'', ed. by D.M. Ginsberg,
World Scientific Publ. Comp., Singapore, 1992, p. 363

\bibitem{Collins} R.T. Collins et al., Phys. Rev. B {\bf 39}, 6571 (1989)

\bibitem{Greco2} A. Greco and R. Zeyher, Phys. Rev. B {\bf 60}, 1296 (1999)

\bibitem{Zeyher3} R. Zeyher and M. Kuli\'c, Phys. Rev. B {\bf 53}, 
2850 (1996)

\bibitem{Andersen} O.K. Andersen, S.Y. Savrasov, O. Jepsen, and A.I.
Liechtenstein, J. of Low Temp. Physics {\bf 105}, 285 (1996) 


%\bibitem{Levin2} Y. Zha, K. Levin, Q. Si, Phys. Rev. B {\bf 47}, 9124 (1993)

%\bibitem{Levin3} Y.-J. Kao, Q. Si, and K. Levin, cond-mat/99(24August99)

\end{thebibliography}
\end{document}